\documentclass[showpacs,preprintnumbers,amsmath,amssymb,APSl,prd,nofootinbib,
superscriptaddress,twocolumn]{revtex4-1}

\usepackage{graphicx,color}
\usepackage{subfig}
\usepackage{amssymb}
\usepackage{hyperref}
\usepackage[utf8]{inputenc}
\usepackage[english]{babel}
\usepackage{epsfig}
\usepackage{wasysym}
\usepackage{color,xcolor}
\usepackage{amsmath}
\usepackage{bm}
\usepackage{epsfig}
\usepackage{amsfonts}
\usepackage{dcolumn}
\usepackage{float}

\usepackage{mathtools, nccmath, hhline}
\usepackage{makecell}
\setcellgapes{3pt}

\usepackage{array, boldline, rotating}

\definecolor{purple}{rgb}{1,0,1}
\definecolor{lime}{HTML}{A6CE39} 

\begin{document}

\title{Observations on the massive particle surface method}

	\author{Ednaldo L. B. Junior} \email{ednaldobarrosjr@gmail.com}
\affiliation{Faculdade de Física, Universidade Federal do Pará, Campus Universitário de Tucuruí, CEP: 68464-000, Tucuruí, Pará, Brazil}

     \author{José Tarciso S. S. Junior}
    \email{tarcisojunior17@gmail.com}
\affiliation{Faculdade de F\'{i}sica, Programa de P\'{o}s-Gradua\c{c}\~{a}o em F\'{i}sica, Universidade Federal do Par\'{a}, 66075-110, Bel\'{e}m, Par\'{a}, Brazill}

	\author{Francisco S. N. Lobo} \email{fslobo@ciencias.ulisboa.pt}
\affiliation{Instituto de Astrof\'{i}sica e Ci\^{e}ncias do Espa\c{c}o, Faculdade de Ci\^{e}ncias da Universidade de Lisboa, Edifício C8, Campo Grande, P-1749-016 Lisbon, Portugal}
\affiliation{Departamento de F\'{i}sica, Faculdade de Ci\^{e}ncias da Universidade de Lisboa, Edif\'{i}cio C8, Campo Grande, P-1749-016 Lisbon, Portugal}

    \author{Manuel E. Rodrigues} \email{esialg@gmail.com}
\affiliation{Faculdade de F\'{i}sica, Programa de P\'{o}s-Gradua\c{c}\~{a}o em F\'{i}sica, Universidade Federal do Par\'{a}, 66075-110, Bel\'{e}m, Par\'{a}, Brazill}
\affiliation{Faculdade de Ci\^{e}ncias Exatas e Tecnologia, Universidade Federal do Par\'{a}, Campus Universit\'{a}rio de Abaetetuba, 68440-000, Abaetetuba, Par\'{a}, Brazil}
    
     \author{Luís F. Dias da Silva} 
        \email{fc53497@alunos.fc.ul.pt}
\affiliation{Instituto de Astrof\'{i}sica e Ci\^{e}ncias do Espa\c{c}o, Faculdade de Ci\^{e}ncias da Universidade de Lisboa, Edifício C8, Campo Grande, P-1749-016 Lisbon, Portugal}

    \author{Henrique A. Vieira} \email{henriquefisica2017@gmail.com}
\affiliation{Faculdade de F\'{i}sica, Programa de P\'{o}s-Gradua\c{c}\~{a}o em F\'{i}sica, Universidade Federal do Par\'{a}, 66075-110, Bel\'{e}m, Par\'{a}, Brazill}


\begin{abstract}

The geodesic method has played a crucial role in understanding the circular orbits generated by compact objects, culminating in the definition of the photon sphere, which was later generalized to a photon surface in arbitrary spacetimes. This new formulation extends the concept of the photon sphere in a broader sense, including dynamical spacetimes, as shown by the Vaidya solution. The photon surface essentially defines the null geodesics, which are originally tangent to the temporal surface, and keeps them confined to this surface. However, this formalism does not cover all classes of particles, and to overcome this limitation, a more comprehensive approach, denoted as the ``massive particle surface'', has been proposed that also accounts for charged massive particles. Indeed, the photon surface concept is recovered when the charge and mass of the particles are zero. In this work, we use these three formalisms to check the consistency of the results for the values of the radius of the photon sphere ($r_{ps}$) and the radius of the ``innermost stable circular orbit" (ISCO) ($r_{\rm ISCO}$) for some gravitational models. In our results, the first model is described by conformal gravity, with the peculiarity that $g_{00}\neq-g_{11}^{-1}$. The second model, i.e. Culetu model, is developed by coupling General Relativity (GR) with nonlinear electrodynamics (NLED), which requires the consideration of the effective metric ($g_{\rm eff}^{\mu\nu}$) for geodesic approaches, for example. 
Under these circumstances, we have found that the value for the radius of the photon sphere ($r_{ps}$) obtained by the massive particle surface formalism in the conformal gravity case does not agree with the values obtained by the geodesic and photon surface formalisms. Similarly, the values for $r_{\rm ISCO}$ differ between the geodesic and the massive particle surface formalisms. In Culetu's model, we found the same values for the radius of the photon sphere $r_{ps}$ when we consider the effective metric in the geodesic and photon surface formalisms. However, when we apply the massive particle surface formalism, we find an inconsistency with the values of the other two formalisms. Finally, we have examined the expressions for $r_{ps}$ and $r_{\rm ISCO}$ for a spherically symmetric and generally static metric arising from the massive particle surface method. We find that the expression for $r_{ps}$, for example, differs from the photon surface method, as does the expression for $r_{\rm ISCO}$, which differs from the geodesic formalism. Moreover, we highlight a significant difference in the two expressions obtained for a static and spherically symmetric metric in general, as they exhibit a dependence on the metric function $-g_{11}=B(r)$, unlike the other expressions for $r_{ps}$ and $r_{\rm ISCO}$ in the other two formalisms.

\end{abstract}

\date{\today}

\maketitle

\section{Introduction}

The photon sphere is a timelike hypersurface in which unstable photons are trapped \cite{Virbhadra:1999nm}. Indeed, the initial understanding of the surface formed by photons, for instance, has enabled insightful studies of the gravitational lensing effect in the strong field region \cite{Bozza:2001xd,Bozza:2002zj,Perlick:2004tq,Virbhadra:2007kw,Virbhadra:2008ws,Perlick:2021aok}. Over the years, this definition has evolved and been generalized, and has led to a new approach,  namely, the photon surface \cite{Claudel:2000yi}, which is described in an arbitrary spacetime. It was shown that, subject to an energy condition, a black hole in any such spacetime must be surrounded by a photon sphere. Conversely, subject to an energy condition, any photon sphere must surround a black hole, a naked singularity or more than a certain amount of matter \cite{Claudel:2000yi}.  

For self-consistency and self-completeness, and to attain a better understanding of the problem, we  present the concept of the photon sphere using the geodesic formalism, where the equations of motion are derived from the Lagrangian approach. We begin our discussion by introducing the geodesic equation and explaining how the photon sphere can be defined by this description. To introduce this concept, consider a static and spherically symmetric metric described by
\begin{equation}
     ds^{2}=A(r)dt^{2}-B(r)dr^{2}-C(r)\left(d\theta^{2}+\sin^2\theta d\phi^{2}\right).\label{Metric}
\end{equation}

As mentioned above, the Lagrangian formalism yields the geodesic equation, given by
\begin{equation}
   {\cal L}=\frac{1}{2}g_{\mu\nu}v^{\mu}v^{\nu},\label{Li}
\end{equation}
where $v^{\mu}=dx^{\mu}/d\lambda$ is the four-velocity and $\lambda$ is the affine parameter.
For the case of photons, we have ${\cal L}=0$. Under this condition and after substituting Eq.\eqref{Metric} into the Lagrangian \eqref{Li}, we obtain the following result
\begin{equation}
   (v^{t})^{2}A(r)-(v^{r})^{2}B(r)-(v^{\phi})^{2}C(r)\sin^{2} \theta -(v^{\theta})^{2}C(r)=0.\label{L0i}
\end{equation}
This expression can be solved for $v^r$, which is given by 
\begin{equation}
    v^r=\sqrt{\frac{(v^{t})^{2}A(r)-(v^{\phi})^{2}C(r)\sin^{2}\theta -(v^{\theta})^{2}C(r)}{B(r)}}. \label{vr}
\end{equation}

Let us now introduce the geodesic equation defined by the Levi-Civita connection, where the radial component is described as follows:
\begin{equation}
\frac{dv^{r}}{d\lambda}=-\Gamma_{\phantom{\alpha}\mu\nu}^{r}v^{\mu}v^{\nu}.\label{eqgeodesici}
\end{equation}
Now, substituting Eq.~\eqref{vr} into the geodesic equation~\eqref{eqgeodesici} yields
\begin{equation}
   \dot{r}^{2}=-\frac{1}{2}\Bigl((v^{\phi})^{2}\sin^{2} \theta +(v^{\theta})^{2}\Bigr)\Bigl(C(r)A'(r)-A(r)C'(r)\Bigr).\label{eqgeodesic2}
\end{equation}

Next, to illustrate how one can determine the radius of the photon sphere using the geodesic equation~\eqref{eqgeodesic2}, for simplicity, consider the Schwarzschild solution represented by the following metric functions
\begin{equation}\label{fm_Schwarzschild}
        A(r)=\frac{1}{B(r)}=1-\frac{2 M}{r},\quad
        C(r)=r^2.
\end{equation}
Substituting these into Eq.~\eqref{eqgeodesic2}, we obtain the following equation for the Schwarzschild case
\begin{equation}
\dot{r}^{2}=(r-3M)\left((v^{\phi})^{2}\sin^{2}\theta +(v^{\theta})^{2}\right),   
\end{equation}
where the overdot denotes the derivative with respect to a given affine parameter.

When $r> 3M$, the expression is positive, which means that light particles follow null geodesics in spacetime and do not fall into the black hole. If, on the other hand, the radial coordinate lies in the range $2M<r<3M$, the expression is negative and describes null geodesics whose particles inevitably fall into the black hole. The photon sphere is defined by the hypersurface at $r_{ps}=3M$ in Schwarzschild spacetime, where light rays describing trajectories that are initially tangential to the photon sphere remain trapped in it. In describing the photon sphere, we generally have two basic properties: (i) the first demands that a geodesic null surface tangent to the photon sphere always remains tangential to it; (ii) the second states that this surface $r_{ps}=3M$ does not evolve over time~\cite{Virbhadra:1999nm}.

Let us now consider the following components for the four-velocity:
    \begin{equation}
        v^t=\frac{{\cal E}}{A(r)}, \qquad
        v^\theta=0, \qquad
        v^\phi=\frac{L}{C(r)}.
        \label{comp_v}
    \end{equation}
If we substitute these components into Eq.~\eqref{vr}, we arrive at the following
\begin{equation}
    v^r=\sqrt{\frac{{\cal E}^{2}C(r)-L^{2}A(r)}{A(r)B(r)C(r)}}. \label{vr2}
\end{equation}
In this way, the result obtained by squaring Eq.~\eqref{vr2} is expressed through the following function
\begin{equation}
    {\cal R}(r)=\frac{{\cal E}^{2}C(r)-L^{2}A(r)}{A(r)B(r)C(r)}. \label{Ri}
\end{equation}
Applying the condition ${\cal R}(r)=0$ to Eq. \eqref{Ri}, we determine the energy 
\begin{equation}
    {\cal E}=L\sqrt{\frac{A(r)}{C(r)}}.\label{Ei}
\end{equation}

Now, by taking the derivative of ${\cal R}(r)$ and substituting the energy \eqref{Ei}, we deduce 
\begin{equation}
   \frac{d{\cal R}(r)}{dr} =\frac{L^2 \Bigl(A(r) C'(r)-C(r) A'(r)\Bigr)}{A(r) B(r) C(r)^2}.
\end{equation}
Applying the condition $d{\cal R}(r)/dr=0$ to the above expression for the Schwarzschild metric functions, we obtain the radius of the photon sphere given by
\begin{equation}
    r_{ps}=3M.\label{rpsi}
\end{equation}

We can verify the stability of closed circular orbits by defining the effective potential
\begin{equation}
    V_{\rm eff}(r)=L^2\frac{A(r)}{C(r)}.
\end{equation}
If we take the second derivation of this potential for the Schwarzschild case, we find that 
\begin{equation}
   \frac{d^2V_{\rm eff}(r)}{dr^2}\Big|_{r=r_{ps}} =-\frac{2 L^2}{81 M^4}.
\end{equation}
Since $d^2V_{\rm eff}(r)/dr^2 < 0$, we have a local maximum, i.e. the photon sphere has unstable circular orbits.  

In the reference study~\cite{Virbhadra:1999nm}, the authors define a photon sphere in a static, spherically symmetric spacetime as a timelike hypersurface, where ${ r= r_{ps} }$, because when $r$ approaches $r_{ps}$, the Einstein angle of curvature of a light beam is unlimitedly large. As mentioned above, the concept of the photon sphere has been improved by the introduction of a new formulation known as the photon surface, developed by Ellis et al.~\cite{Claudel:2000yi}.
Previously, we discussed two properties of the photon sphere, one of which was that these surfaces do not evolve in time. In an attempt to extend this concept, the authors in Ref.~\cite{Claudel:2000yi} generalize the definition of the photon sphere in order to incorporate dynamical spacetimes in the second property of the photon sphere. An example of this approach is the Vaidya~\cite{Vaidya1943,Vaidya1951,Vaidya1953} metric, which can now be included in the description of the photon surface method.

Although the photon surface method is efficient, it has its limitations, for example, it cannot include massive charged particles. Recently, a generalization of the photon surface method was proposed by Gal'tsov et al.  \cite{Kobialko:2022uzj}.
This new approach is defined by a timelike hypersurface with worldlines of charged particles $Q$, mass $m$ and fixed total energy ${\cal E}$, which, when tangent to this hypersurface, remain tangential to it. If we consider $Q=m=0$ in this spacetime, we again obtain the photon surface model proposed in Ref.~\cite{Claudel:2000yi}, i.e. we return to the impact parameter with a fixed value instead of particles with fixed energy, as is the case with the massive particle surface method.

These concepts are extremely important because the surfaces that form near compact objects are directly related to the shadows of black holes~\cite{Falcke:1999pj,Cunha:2015yba,Gralla:2019xty,Perlick:2021aok,Rosa:2022tfv}, gravitational lensing~\cite{Virbhadra:1999nm,Bozza:2002zj,Jusufi:2018jof,Bozza:2012by,Tsukamoto:2020iez,Tsukamoto:2020bjm,Tsukamoto:2021fsz,Zhang:2022nnj,Junior:2023xgl}, and the behavior of the optical appearance for a particular type of radiation emitted by the accretion disk~\cite{Koga:2022dsu,Bogush:2022hop,daSilva:2023jxa}, for example.

An important issue to investigate is that the photon surface method has proven to be highly efficient in numerous solutions of both static \cite{Bozza:2010xqn, Tsukamoto:2016jzh, Perlick2,Stefanov:2010xz} and stationary \cite{Bozza:2002af} black holes, in plasma \cite{Atamurotov:2015nra, Atamurotov, Babar}, as well as dynamic scenarios \cite{Mishra, Solanki, Koga}. Now we inquire, does the massive particle surface method also exhibit the same efficiency and yield consistent results for an extensive range of black hole solutions? Our primary objective here is to demonstrate that there are slight discordances between the massive particle surface method and the geodesic and photon surface methods.

 In this study, we investigate the radius of the photon sphere in different gravitational models using the geodesic, photon surface and massive particle surface methods. The structure of the manuscript is outlined as follows: In Sec.~\ref{sec2}, we briefly introduce the geodesic equations derived from the Lagrangian and discuss how to calculate the radius of the photon sphere and the $r_{\rm ISCO}$. In Sec.~\ref{sec3}, we give a brief definition of the photon surface method and the corresponding equation for spherically symmetric spacetimes. In Sec.~\ref{sec4} we briefly discuss the method of massive particle surfaces and how this description can be applied to a static and spherically symmetric geometry and conclude with an example. In Sec.~\ref{sec5} we present our results using two gravitational models, the first described by conformal gravity and the second by NLED. Finally, in Sec. ~\ref{sec6} we summarize our results.

\section{GEODESICS AND EFFECTIVE POTENTIAL}\label{sec2}

In this section, we determine the equations describing the motion of test particles near a compact object using the Lagrangian formalism. We also discuss the conditions imposed on the geodesic equations to determine, for instance, the radius of the photon sphere. For a deeper understanding, we later present an example of the Reissner-Nordström (RN) solution.

As a first step, consider the static and spherically symmetrical line element~\eqref{Metric}. 

In order to obtain the equations of motion of the particles near the black hole, we use the following Lagrangian formula
\begin{equation}
    {\cal L}=\frac{1}{2}g_{\mu\nu}\dot{x}^{\mu}\dot{x}^{\nu}=m^2,\label{Lagran}
\end{equation}
where the overdot denotes the derivative with respect to an affine parameter  $\lambda$,  $m$ is the mass of the test particle and 
 $\dot{x}^\mu\rightarrow(\dot{t}, \dot{r}, \dot{\theta}, \dot{\varphi} )$ is the four-velocity of the particle.

Using the metric~\eqref{Metric}, we can rewrite Eq. \eqref{Lagran} as
\begin{equation}
    {\cal L}=\frac{1}{2}\left[A\left(r\right)\dot{t}^{2}-B(r)\dot{r}^{2}-C(r)\left(\dot{\theta}{}^{2}+\sin^2\theta\dot{\phi}^{2}\right)\right]=m^2,
\end{equation}
which provides the following canonical moments:
\begin{subequations}
\begin{align}
p_{t}&=\frac{{\cal \partial L}}{\partial\dot{t}}=A\left(r\right)\dot{t},\label{pt}\\
p_{r}&=\frac{{\cal \partial L}}{\partial\dot{r}}=B(r)\dot{r},\label{pr}\\
p_{\theta}&=\frac{{\cal \partial L}}{\partial\dot{\theta}}=C(r)\dot{\theta},\label{ptheta}\\
p_{\phi}&=\frac{{\cal \partial L}}{\partial\dot{\phi}}=C(r)\sin^2\theta\dot{\phi}\label{pphi}.
\end{align}
\end{subequations}

The relevant equations result from the Euler-Lagrange equation
\begin{equation}
    \frac{d}{d \sigma}\left(\frac{{\cal \partial L}}{\partial\dot{x}^{\mu}}\right)-\frac{{\cal \partial L}}{\partial x^{\mu}}=0,
\end{equation}
since the metric~\eqref{Metric} does not depend directly on the coordinates $t$ and $\phi$, it follows that
\begin{subequations}\label{EL}
\begin{align}
{\cal E}&=A\left(r\right)\dot{t},\label{E}\\ 
m^2&=A\left(r\right)\dot{t}^{2}-B(r)\dot{r}^{2}-C(r)\left(\dot{\theta}{}^{2}+\sin^2\theta\dot{\phi}^2\right)\\
L&=C\left(r\right)\dot{\phi},\label{L}
\end{align}
\end{subequations}
where ${\cal E}$ is the energy, $L$ is the angular moment of the particle and we absorb the factor 2 with the transformation $\sqrt{2}m\rightarrow m$.
Due to the spherical symmety, without loss of generality we consider motion in the equatorial plane, i.e. $\theta=\pi/2$ and $\dot{\theta}=0$, which implies $p_\theta=0$.
\par
Using these results, we thus obtain the geodesic equation for test particles:
\begin{equation}
\dot{r}^{2}=\frac{1}{A(r)B(r)}\left(\frac{1}{b^2}-V_{\rm eff}(r)-m^2 A(r)\right)\,,
\label{eq_geo} 
\end{equation}
where $b$ is the impact parameter defined by
\begin{equation}
b \equiv \frac{L}{{\cal E}}. 
\end{equation}
and the effective potential represented in Eq.~\eqref{eq_geo} is defined as
\begin{equation}
    V_{\rm eff}(r)=\frac{A(r)}{C(r)}.\label{V}
\end{equation}

By considering geodesics with a constant radial coordinate $r$, the photon sphere is obtained by the imposition of two constraints, given by
\begin{equation}\label{CC}
    \dot{r}=0,\qquad    \ddot{r}=0.
\end{equation}
These are the boundary conditions that allow us to determine the radius of the photon sphere, which we denote by $r_{ps}$, and the critical impact parameter $b_c$. If the value of $b$ is larger than $b_c$, the particles can escape the gravitational influence of the black hole. If $b$ is significantly smaller than $b_c$, the particles fall into the black hole. In the critical case, i.e., when $b=b_c$, the particles with this value are confined to an unstable orbit known as the photon sphere. 

Their explicit forms are described below
\begin{equation}
    r_{ps}=2\frac{A(r_{ps})}{A^\prime(r_{ps})},\label{rps}
\end{equation}
where the prime denotes a derivative with respect to the coordinate $r$. 

The radius of the black hole shadow with respect to an observer located at distance $r_O$ is given by~\cite{Vagnozzi:2022moj}
\begin{equation}
    r_{sh}=r_{ps}\sqrt{\frac{A(r_O)}{A(r_{ps})}},
\end{equation}
where  $r_O$ denotes the location of the observer.  

The next step is to find the radius that delineates between the last stable orbits and the unstable orbits using the geodesic formalism we have been employing up to this point. This orbit is known as the ``innermost stable circular orbit'' (ISCO). To determine this radius, we perform a redefinition in Eq.~\eqref{eq_geo}, setting 
\begin{equation}
    \dot{r}^2 \longleftrightarrow{\cal R}(r).\label{R}
\end{equation}
In this way, we can use the above redefinition, Eq.~\eqref{R}, to express Eq.~\eqref{eq_geo} as follows:
\begin{equation}
    {\cal R}\left(r\right)=\frac{1}{A(r)B(r)}\left[{\cal E}^{2}-\frac{A(r)}{C\left(r\right)}\left(m^{2}C\left(r\right)+L^{2}\right)\right].\label{R2}
\end{equation}

With this subtlety in mind, we express the conditions~\eqref{CC} as follows:
\begin{subequations}\label{CC2}
\begin{align}
    {\cal R}(r)&=0,\label{R0}\\
    \frac{d{\cal R} (r)}{dr} &=0.\label{dR0}
\end{align}
\end{subequations}
From these expressions we may obtain, for example, the explicit forms of the parameters ${\cal E}$ and $L$. Having obtained these quantities, we substitute their values into ${\cal R}(r)$ and determine $r_{\rm ISCO}$ using the condition 
\begin{equation}
    \frac{d^2{\cal R} (r)}{dr^2}=0.\label{CC3}
\end{equation}
This approach will become clearer in the examples presented later.

Thus, if we apply condition~\eqref{R0} to Eq.~\eqref{R2}, we find that the energy takes the form:
\begin{equation}
   {\cal E}= \sqrt{A(r)\left(m^{2}+\frac{L^{2}}{C(r)}\right)}\,.
   \label{E_geod}
\end{equation}
On the other hand, if we apply condition~\eqref{dR0} to Eq.~\eqref{R2}, we find that the angular momentum takes the following form:
\begin{equation}
   L= \frac{mC\left(r\right)\sqrt{A'(r)}}{\sqrt{A(r)C^{\prime}\left(r\right)-C\left(r\right)A'(r)}}\,.
   \label{L_geod}
\end{equation}

To determine the radius $r_{\rm ISCO}$, we start by developing the second derivative of ${\cal R}(r)$, as shown in Eq.~\eqref{R2}. In the expression resulting from this procedure, we insert the expressions for energy and angular momentum defined in Eqs. \eqref{E_geod} and \eqref{E_geod}, respectively. In this way we obtain
\begin{widetext}
\begin{equation}
   \frac{d^2{\cal R} (r)}{dr^2}= \frac{m^{2}\left[C\left(r\right)\Bigl(-A(r)A''(r)C^{\prime}\left(r\right)+A(r)A'(r)C^{\prime\prime}\left(r\right)+2A'(r)^{2}C^{\prime}\left(r\right)\Bigr)-2A(r)A'(r)\bigl(C^{\prime}\left(r\right)\bigr)^{2}\right]}{A(r)B(r)C\left(r\right)\Bigl(A(r)C^{\prime}\left(r\right)-C\left(r\right)A'(r)\Bigr)}.\label{dR2_geod}
\end{equation}
\end{widetext}
So if we substitute the metric function $A(r)$ and $C(r)$ into Eq.~\eqref{dR2_geod} and then apply condition~\eqref{CC3}, we find an expression that allows us to determine the $r_{\rm ISCO}$ by solving in terms of $r$.

\subsection{Specific example: Reissner-Nordström space-time}

As a specific example, consider the Reissner-Nordström asymptotically flat spacetime, which is a static and spherically symmetric metric around a body with mass $M$ and electric charge $q$, and is characterized by the following metric functions: 
\begin{subequations}\label{fm_RN}
    \begin{align}
        A(r)=\frac{1}{B(r)}&=1-\frac{2 M}{r}+\left(\frac{q}{r}\right)^2,\label{A_RN}\\
        C(r)&=r^2.
    \end{align}
\end{subequations}

The event horizons $r_H$ are found for $A(r)=0$, which yields:
\begin{equation}
    r_H=M\pm \sqrt{M^2-q^2}.
\end{equation}

From Eq.~\eqref{rps} we determine the radius of the photon sphere, where the explicit form is given by
\begin{equation}
    r_{ps}=\frac{1}{2} \left(\sqrt{9 M^2-8 q^2}+3 M\right).
    \label{rps_RN}
\end{equation}
If we assume that the charge is zero, $q=0$, the above expression reduces to the radius of the Schwarzschild photon sphere, thus $r_{ps}=3M$

Applying the metric functions~\eqref{fm_RN} to Eq.~\eqref{eq_geo}, we find that Eq.~\eqref{R2} in this case has the following form:
\begin{equation}
    {\cal R}(r)={\cal E}^2-\frac{\left(L^2+M^2 r^2\right) \left[r (r-2 M)+q^2\right]}{r^4}.\label{R_RN}
\end{equation}

Applying the conditions~\eqref{CC2} to Eq. ~\eqref{R_RN}, we obtain the expressions for ${\cal E}$ and $L$ as follows:
\begin{subequations}\label{EL_RN}
    \begin{align}
       {\cal E} =&\frac{ M \Big[r (r-2 M)+q^2\Big]}{\sqrt{r^2 \Big[r (r-3 M)+2 q^2\Big]}},\\
       L=&\frac{ M r \sqrt{M r-q^2}}{\sqrt{r (r-3 M)+2 q^2}}.\label{L_RN}
    \end{align}
\end{subequations}

To determine the value of $r_{\rm ISCO}$, we first insert the above expressions, Eqs.~\eqref{EL_RN}, into the Eq.~\eqref{R_RN}. For example, we can calculate the second derivative with respect to $r$, which yields
\begin{equation}
    \frac{d^{2}{\cal R}}{dr^{2}}=\frac{2 M^2 \left[-9 M q^2 r+M r^2 (6 M-r)+4 q^4\right]}{r^4 \left[r (r-3 M)+2 q^2\right]}.\label{dr2}
\end{equation}

\begin{figure}[h]
\centering
\includegraphics[scale=0.4]{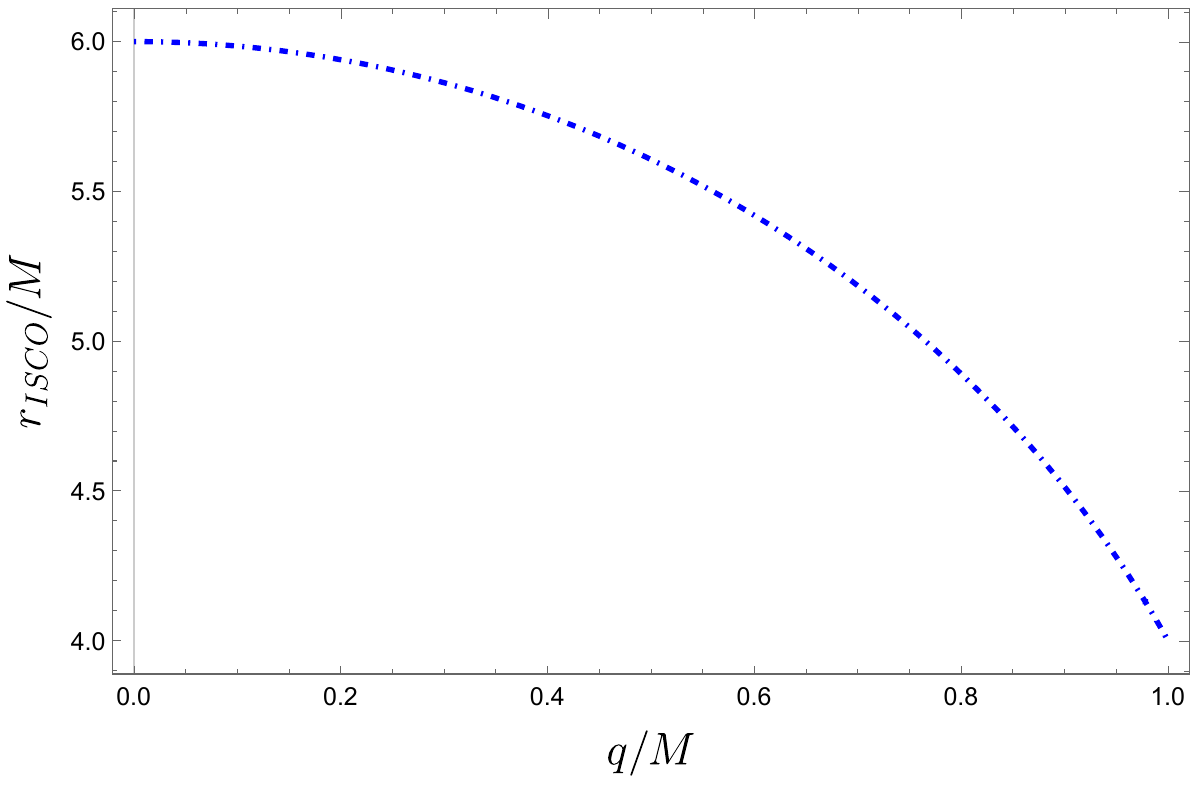}
\caption{Plot of $r_{\rm ISCO}/M$ with respect to the $q/M$, representing the real roots obtained from  $r_{\rm ISCO}$.}
\label{rISCO_RN}
\end{figure}

Thus, if we apply the condition~\eqref{CC3} to Eq.~\eqref{dr2}, we find that the value of $r_{\rm ISCO}$ for this model is described by three roots.
For $0 < q\leq M$ there is only one real root, which we have represented in Fig.~\ref{rISCO_RN}. Moreover, we can see in this graph that we get the Schwarzschild $r_{\rm ISCO}$ if we choose $q=0$, i.e. $r_{\rm ISCO}=6M$.

Without a loss of generality, we can choose a value for the charge to represent the $r_{\rm ISCO}$ in terms of mass, for instance, if $q = 0.2$, we get
\begin{equation}
    r_{\rm ISCO}=5.939 M.\label{ri_RN}
\end{equation}

\section{Photon surface}\label{sec3}

We will show how to express the photon sphere with a more comprehensive formalism than the geodesic formalism. We will not go into all the mathematical approaches leading to one of the main equations of this proposal, but we refer the reader to \cite{Claudel:2000yi} for more details. As a starting point, it is useful to define a manifold ${\cal M}$ whose connection is described by the metric tensor $g_{\mu\nu}$, and ${\cal S}$ denotes the hypersurface. In the formulation proposed by Ellis et al. in~\cite{Claudel:2000yi}, they define the photon surface by the null geodesics which, when tangent to the hypersurface, remain trapped on the photon surface, i.e. on ${\cal S}$. Formally, the authors formulate this definition as follows.\\

\textbf{Definition 2.1.} A photon surface of $({\cal M}, g)$ is an immersed, nowhere-spacelike
hypersurface ${\cal S} $ of $({\cal M}, g)$ such that, for every point $p \in {\cal S}$ and every null vector ${\bf k} \in T_p
{\cal S}$, there exists a null geodesic $\gamma : (-\epsilon, \epsilon) \to {\cal M} $ of $({\cal M}, g)$ such that $\dot{\gamma}(0) =\bf{k}$, $\mid\gamma\mid\subset {\cal S}$. \\

Using the metric~\eqref{Metric}, the photon sphere for a spherically symmetric static spacetime can therefore be determined as follows
\begin{equation}
A(r)C^\prime(r)=A^\prime(r)C(r).\label{phot_spher}
\end{equation}

\section{Massive particle surfaces}\label{sec4}

In this section, we will briefly introduce the formalism developed by Galt'sov et al~\cite{Kobialko:2022uzj}, which is used to define the massive particle surface method. This concept extends the previously discussed understanding of ``photon surfaces" to the case of charged particles with mass.
In constructing this formalism, the authors make use of the symmetry of Killing vectors. As a starting point, it is useful to define a Lorentz manifold $\cal{M}$ of dimension $n\geq4$ in which the connection of the covariant derivative is described by the Christoffel symbol, i.e. in terms of the metric $g_{\mu\nu}$. In addition, some electromagnetic quantities are considered, including the Maxwell-Faraday tensor, which is defined as $F_{\mu \nu} = \partial_\mu A_\nu -\partial_\nu A_\mu$, where ${A_\mu}$ is the electromagnetic vector potential, and $F=\frac{1}{4}F^{\mu\nu}F_{\mu\nu}$, is the electromagnetic scalar.

A test particle with charge $Q$ and mass $m$ therefore describes a world line $\gamma^\mu$ under the following conditions
 \begin{subequations}
\begin{align}
\dot{\gamma}^{\mu}\nabla_{\mu}\dot{\gamma}^{\nu}& = Q F_{\phantom{\nu}\sigma}^{\nu}\dot{\gamma}^{\sigma},\\
\dot{\gamma}^{\mu}\dot{\gamma}_{\mu}&=-m^{2}\,,
\end{align}
\end{subequations}
with the four-velocity of the particle defined as:
\begin{equation}
    \dot{\gamma}^{\mu}=\frac{d\gamma^\mu}{d\lambda}.
\end{equation}

Let us now consider the context of symmetry in spacetime with respect to the Killing vectors $k^\mu$, which apply to static and stationary geometries. In this scenario, the metric fulfills the Lie derivation
    \begin{equation}
        {\cal L} _k g_{\mu\nu}=\nabla_{(\mu}g_{\nu)}=0,
    \end{equation}
and the electromagnetic potential fulfils the following condition
\begin{equation}
{\cal L}_{k}A_{\mu}=k^{\sigma}\nabla_{\sigma}A_{\mu}+\nabla_{\mu}k^{\sigma}A_{\sigma}=0.
\end{equation}

The total energy of the particle is thus preserved along the world lines, and its expression has the following form
\begin{equation}
    {\cal E}=-k_{\mu}\left(\dot{\gamma}^{\mu}+ Q A^{\mu}\right).
\end{equation}
In terms of the kinetic and potential energy, ${\cal E}$ can be expressed as follows
\begin{subequations}
    \begin{align}
      {\cal E}_p  &\equiv -q k_\mu A^\mu\,, \\
      {\cal E}_k  &\equiv{\cal E}-{\cal E}_p=-k_{\mu}v^{\mu}\,.
    \end{align}
\end{subequations}

Let us now show some relevant quantities to determine the particles with total energy $\cal{E}$, mass $m$ and charge $Q$, which are fixed on the hypersurface $\cal{S}$.
For example, the metric of the induced hypersurface is expressed as follows
\begin{equation}
    h_{\mu\nu}=g_{\mu\nu}-n_{\mu}n_{\nu},
\end{equation}
and the second symmetrical fundamental form \begin{equation}
    \chi_{\mu\nu}\equiv h_{\phantom{\lambda}\mu}^{\lambda}h_{\phantom{\sigma}\nu}^{\sigma}\nabla_{\lambda}n_{\sigma},
\end{equation}
where $n^{\sigma}$ is the unit space normal vector of the temporal hypersurface $\cal{S}$ of dimension $n-1$ and the projector operator is defined as $h_{\phantom{\mu}\nu}^{\mu}=\delta_{\phantom{\mu}\nu}^{\mu}-n^{\mu}n_{\nu}$.

The projection of the Killing vector onto the hypersurface is
\begin{subequations}
    \begin{align}
        k^{\mu}&=\kappa^{\mu}+k\perp n^{\mu},\\
        \kappa^{\mu}n_{\mu}&=0.
    \end{align}
\end{subequations}

Based on the above quantities, the authors in Ref.~\cite{Kobialko:2022uzj} cover the concept of a photon sphere as defined in~\cite{Claudel:2000yi}. The generalization of this concept is called ``massive particle surfaces". In this new definition, we consider a particle with mass $m$, electric charge $Q$ and fixed total energy ${\cal E}$ that is tangent to the massive particle surfaces ${\cal S}_{{\cal E}}$, provided that the world line of the particle is initially tangent to ${\cal S}_{{\cal E}}$.  This concept is defined in mathematical terms as follows. \\

\textbf{Definition} - A massive particle surface in ${\cal M}$ is an immersed, timelike, nowhere orthogonal to Killing vector $k^\mu$ hypersurface ${\cal S} _{\cal E}$ of ${\cal M}$ such that, for every point $p \in {\cal S} _{\cal E}$ and every vector $v^{\mu}\mid_{p}\in T_{p}{\cal S}_{{\cal E}}$ such that  $v^{\mu}\kappa_{\mu}\mid_{p}={\cal E}_{k}\mid_{p}$ and $v^{\mu}v_{\mu}\mid_{p}=m^{2}$, there exists a worldline $\gamma$ of ${\cal M}$ for a particle with mass $m$, electric charge $Q$ and total energy $\varepsilon$
such that $\dot{\gamma}^{\mu}(0)=v^{\mu}\mid_{p}$ and $\gamma\in{\cal S}_{\cal E}$.\\

Note that if we assume that the particle has the mass $m=0$, the charge $Q=0$ and the constraints on the total energy, we find the definition of photon surfaces from \cite{Claudel:2000yi}.

Finally, we will not discuss all the mathematical steps of this formulation in detail; we refer the reader to the original work~\cite{Kobialko:2022uzj} for specific details. In the following, we present the main expression of this description, which will later allow us to calculate the radius of the photon sphere for some gravitational models. The total energy of the test particles is thus described by
\begin{eqnarray}
    {\cal E}_{\pm}=\pm m\sqrt{\kappa^{2}\frac{\chi_{\lambda}}{K}+\frac{{\cal F}^{2}\left(n-2\right)^{2}Q^{2}}{4m^{2}K^{2}}}
	\nonumber \\    
    +\frac{{\cal F}\left(n-2\right)Q}{2K} - Q k_{\mu}A^{\mu},
\end{eqnarray}
with the following definitions
\begin{subequations}
\begin{align}
    \chi_{\lambda}&=\frac{\alpha(n-2){\cal E}_{k}^{2}}{m^{2}+\frac{{\cal E}_{k}^{2}}{\kappa^{2}}}\label{chi},
\\
    K&=2 \chi -3 \chi _\lambda,\label{K}
\\
    {\cal F}_{\mu\nu }&=\frac{1}{2}n^{\sigma}F_{\sigma(\mu}\kappa_{\nu)},
\\
 {\cal F}&=n^{\sigma}F_{\sigma\rho}\kappa^{\rho}.
\end{align}
\end{subequations}
where $\chi=\chi^{\alpha}_{\;\;\alpha}$.

\subsection{Static symmetry}

We will now show how some quantities are described in a four-dimensional spacetime with static and spherical symmetry. At the beginning of this example, let us consider the line element with the form of
\begin{equation}
    ds^{2}=A(r) dt^{2}-B(r) dr^{2}-C(r) d\theta^{2}-D(r,\theta) d\phi^{2},
\end{equation}

The second basic form can therefore be expressed as follows
\begin{eqnarray}
\chi_{\rho\sigma}dx^\rho dx^\sigma=\frac{1}{2\sqrt{B(r)}}\Bigl(\partial_{r}A(r) dt^{2}
	\nonumber \\
-\partial_{r}C(r) d\theta^{2}
-\partial_{r}D(r,\theta) d\phi^{2}\Bigl),
\label{seg_form}
\end{eqnarray}
where the trace takes the form
\begin{equation}
\chi=\frac{\partial_{r}\ln\Bigl[A(r)C(r)D(r,\theta)\Bigr]}{2\sqrt{B(r)}},\label{chi2}
\end{equation}
and the Killing vector is chosen as follows $k^\sigma\partial_\sigma=\partial_t$. 
The  mean curvature (\ref{chi}) is given by:
\begin{equation}
\chi_\lambda=2\sqrt{\frac{A(r)}{C(r)}}.
\label{chi_afim}
\end{equation}

Since we are only dealing with electrically charged cases, we only have the following expression
\begin{equation}
{\cal F}_{\rho\sigma}dx^{\rho}dx^{\sigma}=-\frac{A(r)}{\sqrt{B(r)}}\partial_{r}A_{t}dtdt,
\end{equation}
and the contraction of the above quantity yields
\begin{equation}
{\cal F}=\frac{\partial_{r}A_{t}}{\sqrt{B(r)}}.\label{F_scalar}
\end{equation}

\subsection{Reissner-Nordström}

 Similarly to the example we provided with the geodesic formalism, we will use the Reissner-Nordström metric to illustrate obtaining the $r_{\rm ISCO}$ through the massive particle surface formalism.

As a starting point, we calculate the electromagnetic tensor for this description, which is given by
$ F_{10}=-2 q/r^2$.
If we integrate this result with respect to the radial coordinate $r$, we obtain the electromagnetic potential $ A_0=2 q/r$.

This makes it possible to determine ${\cal F}$ for the current model with Eq.~\eqref{F_scalar}. In this particular situation, the following steps are required: 
\begin{equation}
   {\cal F} =-\frac{2 q \sqrt{r (r-2 M)+q^2}}{ r^3}.
\end{equation}

The description of the scalar of the second fundamental form is expressed by the metric functions~\eqref{fm_RN} as follows:
\begin{eqnarray}
   && \chi=\frac{r(2r-3M)+q^{2}}{r^{3}\sqrt{-2Mr+q^{2}+r^{2}}}.\label{chi_RN}
\end{eqnarray}
And mean curvature~\eqref{chi_afim} is now 
\begin{equation}
   \chi_\lambda=\frac{2 }{r} \sqrt{-\frac{2 M}{r}+\frac{q^2}{r^2}+1}.\label{chi_afim_RN}
\end{equation}

Using results~\eqref{chi_RN} and~\eqref{chi_afim_RN}, we can therefore determine the form of $K$ for this model from Eq.~\eqref{K}. Its form is thus described by:
\begin{align}
    &K=-\frac{2}{r^{2}}\left(-2Mr+q^{2}+r^{2}\right)^{-1/2}\left[r(r-3M)+2q^{2}\right].
\end{align}

With these quantities, the energy of the test particle for this case is therefore given by
\begin{eqnarray}
    {\cal E}&=&\frac{m\left[r(r-2M)+q^{2}\right]\sqrt{Q^{2}\left(\frac{q^{2}}{\kappa^{4}m^{2}}+2\right)+r(r-3M)}}{r\left[r(r-3M)+2q^{2}\right]}
	\nonumber \\    
&&   +\frac{qQ\left[r(r-2M)+q^{2}\right]}{\kappa^{2}mr\left[r(r-3M)+2q^{2}\right]}-\frac{2qQ}{\kappa^{2}r}.\label{E2_RN}
\end{eqnarray}
We can determine the radius of the photon sphere using Eq.~\eqref{E2_RN} under the condition $1/{\cal E}^2=0$, ($Q=0$). Remarkably, the result obtained is equivalent to the formalism of the geodesic method
\begin{equation}
    r_{ps}=\frac{1}{2} \left(\sqrt{9 M^2-8 q^2}+3 M\right).\label{rps2_RN}
\end{equation}

We can calculate the $r_{\rm ISCO}$ by using the energy equation for the test particle from Eq.~\eqref{E2_RN}, under the condition $d {\cal E}/dr=0$. The value of the radius $r_{\rm ISCO}$ in this case, if we choose the value of the charge $q=0.2$, is:
\begin{equation}
    r_{\rm ISCO}=5.939 M.\label{ri2_RN}
\end{equation}
Note that the results for the radius of the photon sphere~\eqref{rps2_RN} and the $r_{\rm ISCO}$~\eqref{ri2_RN} agree with the equivalent quantities obtained in the geodesic formalism according to the Eqs.~\eqref{rps_RN} and ~\eqref{ri_RN}.

\section{Results}\label{sec5}

In this section, we discuss the results obtained by choosing two specific metrics that we have applied in the formalisms briefly presented in the previous sections. First, we will deal with a metric derived from conformal gravity, which is characterized by a peculiarity in its metric functions~$g_{00}\neq -g_{11}^{-1}$. Next, we study the Culetu model which results from the coupling of NLED with GR \cite{Culetu:2013fsa,Culetu:2014lca,Simpson:2019mud}. Since this model is described by NLED, it is necessary to use an effective metric to obtain a more appropriate interpretation of the particle motion. For example, we obtain the radius of the photon sphere for these two formulations.

\subsection{Conformal gravity}

In this section, we briefly discuss a conformal gravity solution. It is well known that GR has solutions that involve singularities in spacetime, and faced with this challenge, various proposals have been explored in the literature to overcome this situation. In particular, Weyl's conformal symmetry, or conformal gravity, is a promising proposal to address and solve the problem of spacetime singularities in Einstein's theory of gravity \cite{Mannheim:2011ds}.

In \cite{Bambi:2016wdn}, the authors used the general invariant theory under conformal Weyl transformations to study the regularity of spacetime. In this approach, they use conformal symmetry, more precisely, a conformal scaling of the metric, so that the new metric becomes singularity-free.


For our discussion, we briefly consider a special case of conformal gravity, which we present below.
Einstein's conformal gravity is a second-order conformally invariant formulation described in a multidimensional spacetime $N$.
In this description, the metric is replaced by the introduction of an auxiliary dilaton field~$\varphi$ and by the use of the metric $\bar{g}_{\mu\nu}$, as shown below
\begin{equation}
    g_{\mu\nu}=(\varphi\,\kappa_{N})^{\frac{4}{N-2}}\bar{g}_{\mu\nu}\,.\label{m_CG}
\end{equation}

\subsubsection{Schwarzschild metric in conformal gravity}

Now, we express the Schwarzschild metric $\overset{S}{g}_{\mu\nu}$ in terms of $\varphi$ and $\bar{g}_{\mu\nu}$, given by
\begin{equation}
    \overset{S}{g}_{\mu\nu}=(\varphi\kappa_{N})^{\frac{4}{N-2}}\bar{g}_{\mu\nu}\,.
    \label{m_CG_S}
\end{equation}

Thus, a Weyl rescaling is performed that keeps this approach invariant. This scaling is applied to both the metric and the scalar field, with an arbitrary local parameter $\Omega$, as described by 
\begin{align}
\bar{g}_{\mu\nu}^{*}=\Omega^{2}\,
\bar{g}_{\mu\nu},
\\
\quad\varphi^{*}=\Omega^{\frac{2-D}{2}}\,\varphi.
\end{align}
so that Eq.~\eqref{m_CG_S} takes the form
\begin{equation}
    \overset{S}{g}_{\mu\nu}=(\varphi^*\kappa_{N})^{\frac{4}{N-2}}\bar{g}^*_{\mu\nu}\,,\label{m_CG_S2}
\end{equation}
and the Schwarzschild spacetime is recovered when 
$\varphi=k_D^{-1}$ and $\Omega=1$.

This conformal rescaling imposes no limits on the number of exact solutions equivalent to the Schwarzschild metric. Next, we will show a specific example of an exact black hole solution obtained through this formalism. For simplicity, we consider $D=4$, as described in Ref.~\cite{Bambi:2016wdn}. Thus, the metric is given by:
\begin{align}
& ds^{*2}\equiv\bar{g}_{\mu\nu}^{*}dx^{\mu}dx^{\nu}=S(r)\bar{g}_{\mu\nu}dx^{\mu}dx^{\nu}
   \nonumber \\
    &=S(r)\left[\left(1-\frac{2M}{r}\right)dt^{2}+\left(1-\frac{2M}{r}\right)^{-1}dr^{2}+r^{2}d\Omega^{2}\right],
\end{align}
with
\begin{equation}
\varphi^{*}=S(r)^{-1/2}\kappa_{4}^{-1}\,,
\end{equation}
where the scalar factor is described by   
\begin{equation}
    S(r) = \frac{1}{r^2} \left( \frac{l^4}{r^2}+r^2 \right),
    \end{equation}
and $l$ is a length scale.  

In this way we obtain the following metric functions:
\begin{subequations}\label{fm_CG}
    \begin{align}
        A(r)=&\frac{1}{r^{2}}\left(\frac{l^{4}}{r^{2}}+r^{2}\right)\left(1-\frac{2M}{r}\right)\,,
        \label{A_CG} \\
        B(r)=&\frac{1}{r^{2}}\left(\frac{l^{4}}{r^{2}}+r^{2}\right)\left(1-\frac{2M}{r}\right)^{-1}\,,
        \\
        C(r)&=\left(\frac{l^{4}}{r^{2}}+r^{2}\right)\,.
    \end{align}
\end{subequations}

\subsubsection{Geodesics in conformal gravity} 

The radius of the horizon in this case, from the expression $A(r)=0$, is identical to that of Schwarzschild $ r_H=2  M$.
%
From Eq.~\eqref{rps}, we obtain  the radius of the photon sphere
\begin{equation}
    r_{ps}=3 M.\label{rps_CG}
\end{equation}

The expression for ${\cal R}(r)$ can be determined by applying the metric functions~\eqref{fm_RN} to Eq.~\eqref{eq_geo}, and is given by:
%
\begin{eqnarray}
    {\cal R}(r)=\Big\{r^{7}\left[{\cal E}^{2}r+m^{2}(2M-r)\right]+L^{2}r^{5}(2M-r)
	\nonumber \\    
    +l^{4}m^{2}r^{3}(2M-r)\Big\}\Big/\left(l^{4}+r^{4}\right)^{2}.
    \label{R_CG}
\end{eqnarray}

The conditions~\eqref{CC2} are used to obtain ${\cal E}$ and $L$, which take the following form
\begin{subequations}\label{EL_CG}
    \begin{align}
       {\cal E} =&\frac{ m \sqrt{r^4-l^4} (r-2 M)}{\sqrt{r^5 (r-3 M)}},\\
       L=&\frac{ m \sqrt{l^4 (5 M-2 r)+M r^4}}{r \sqrt{r-3 M}},
       \label{L_conf}
    \end{align}
\end{subequations}
respectively.

Substituting Eqs.~\eqref{EL_CG} into Eq.~\eqref{R_CG}, and taking their second derivative with respect to $r$, we arrive at
\begin{equation}
    \frac{d^{2}{\cal R}}{dr^{2}}=\frac{2m^{2}r\left[l^{4}\left(30M^{2}-21Mr+4r^{2}\right)+Mr^{4}(r-6M)\right]}{\left(l^{4}+r^{4}\right)^{2}(3M-r)}.\label{dr2_CG}
\end{equation}

Let us now apply the condition described in Eq.\eqref{CC3} to Eq.~\eqref{dr2_CG} in order to find the $r_{\rm ISCO}$ for this model. Without a loss of generality, we can choose a value for $l$ to represent $r_{\rm ISCO}$ in terms of mass. So if we choose $l=0.9$, the value of $r_{\rm ISCO}$ is given by:
\begin{align}
    r_{\rm ISCO}&=5.97564 M.\label{risco_CG}
    \end{align}

\subsubsection{Photon surface in conformal gravity} 

By using the metric functions of conformal gravity~\eqref{fm_CG} in Eq.~\eqref{phot_spher} with the choice of $l=0.9$, we determine the radius of the photon sphere, which is given by
\begin{equation}
    r_{ps}=3M.\label{rps2_CG}
\end{equation}
Note that this is the same result as in the Schwarzschild case, which we have already obtained using the geodesic method, Eq.~\eqref{rps_CG}.

\subsubsection{ Massive particle surfaces  in conformal gravity} 

Now we use the metric functions described by Eqs.~\eqref{fm_CG} to express the quantities needed to determine the radius of the photon sphere by the massive particle surfaces formalism in conformal gravity. The first of these quantities is the scalar of the second fundamental form~\eqref{chi}, which is now given by:
\begin{equation}
    \chi=\frac{\sqrt{r}\left[l^{4}(9M-4r)+r^{4}(2r-3M)\right]}{\left(l^{4}+r^{4}\right)^{3/2}\sqrt{r-2M}}.\label{chi_CG}
\end{equation}
The average curvature~\eqref{chi_afim}, in turn, is expressed as  
\begin{equation}
   \chi_\lambda=\frac{2}{r^{2}}\sqrt{\left(\frac{l^{4}}{r^{2}}+r^{2}\right)\left(1-\frac{2M}{r}\right)}.\label{chi2_CG}
\end{equation}

With the help of these quantities, given by Eqs.~\eqref{chi_CG} and~\eqref{chi2_CG}, we obtain $K$, Eq.~\eqref{K}, which takes the form
\begin{eqnarray}
    K=\frac{2}{r^{7/2}}\Bigg\{ \frac{r^{4}\left[l^{4}(9M-4r)+r^{4}(2r-3M)\right]}{\left(l^{4}+r^{4}\right)^{3/2}\sqrt{r-2M}}
	\nonumber \\    
    -3\sqrt{\left(l^{4}+r^{4}\right)(r-2M)}\Big\}. 
    \label{K_CG}
\end{eqnarray}

For this case, in which ${\cal F}=0$, we can therefore only determine the energy of the test particle using the results of Eq.~\eqref{chi2_CG} and Eq.~\eqref{K_CG}, which is takes the form
\begin{eqnarray}
    {\cal E}&=&\frac{\left[\left(l^{4}+r^{4}\right)(r-2M)\right]^{3/2}}{r^{5}}\Big[3\sqrt{\left(l^{4}+r^{4}\right)(r-2M)}
	\nonumber \\    
  &&  -\frac{r^{4}\left[l^{4}(9M-4r)+r^{4}(2r-3M)\right]}{\left(l^{4}+r^{4}\right)^{3/2}\sqrt{r-2M}}\Bigg]^{-1}\,.
    \label{E_CG}
\end{eqnarray}

We can calculate the radius of the photon sphere for this method using Eq.~\eqref{E_CG}, but for this we use $1/{\cal E}^2=0$. For comparison with the result of the geodesic method~\eqref{rps_CG}, we choose the value of the conformal parameter $l=0.9$ without loss of generality. In this way we obtain
\begin{equation}
    r_{ps}=2.92584 M.\label{rps3_CG}
\end{equation}

We also calculate the derivative of ${\cal E}$ with respect to $r$, but we will not express it here as it is lengthy. We determine the radius using Eq.~\eqref{E_CG} under the condition $d{\cal E}/dr=0$. Again, we use the value for the conformal parameter $l=0.9$. We therefore obtain for this case that the radius is equal:
\begin{equation}
    r_{\rm ISCO}=5.70275 M.\label{risco2_CG}
\end{equation}

We thus find that the result for the radius of the photon sphere in the geodesic and photon surface formalisms, Eqs~\eqref{rps_CG} and~\eqref{rps2_CG} respectively, does not agree with the result we obtained for the radius of the photon sphere using the massive particle surface method, Eq.~\eqref{rps3_CG}.
If we use this conformal gravity model to calculate ($r_{\rm ISCO}$) with both formalisms, the geodesic method and the massive particle surface method, we also find a discrepancy in the results, as can be seen in Eqs. \eqref{risco_CG} and \eqref{risco2_CG}.

\subsection{Culetu model}

The next model we deal with is a proposal with static and spherically symmetric symmetry for a regular black hole, i.e. it provides solutions without singularities. This model is known as the Culetu model and it is derived by taking into account that GR  is coupled with NLED \cite{Culetu:2013fsa,Culetu:2014lca}. Moreover, this solution is asymptotic of the Reissner-Nordström type.

Let us now briefly introduce the description of GR  coupled with NLED. The action that provides the description for Culetu's model is given by:
\begin{equation}
   S=\int  d^4x \sqrt{-g} \Bigl[R+\kappa_G^2{\cal L}(F)\Bigr],\label{action}
\end{equation}
where $g$ is the determinant of the metric tensor, $R$ is the Ricci scalar, $\kappa_G^2=8\pi$, ${\cal L}(F)$ is the Lagrangian density of NLED, and $F=\frac{1}{4}F^{\mu\nu}F_{\mu\nu}$
is the electromagnetic scalar and $F_{\mu\nu}$ denotes the Faraday-Maxwell tensor which is represented by the electromagnetic potential vector $F_{\mu \nu} = \partial_\mu A_\nu -\partial_\nu A_\mu$.

Varying the action \eqref{action} with respect to the metric, we find the equations of motion $R_{\mu\nu}-\frac{1}{2}g_{\mu\nu}R=\kappa^2T_{\mu\nu}$,
where the energy-momentum tensor of NLED is given by:
\begin{equation}
    T_{\mu\nu}=\frac{2}{\kappa_{G}^{2}}\left(g_{\mu\nu}{\cal L}(F)-{\cal L}_{F}F_{\mu\alpha}F_{\nu}^{\phantom{\nu}\alpha}\right).
\end{equation}

Varying the action \eqref{action} with respect to $F_{\mu\nu}$, we find
\begin{equation}
    \nabla_\mu ({\cal L}_F F^{\mu\nu})=\frac{1}{\sqrt{-g}} \partial_\mu (\sqrt{-g} {\cal L}_F F^{\mu\nu})=0,
\end{equation}
where ${\cal L}_F=\partial {\cal L} (F)/\partial F$.

The component of $F_{\mu\nu}$ for magnetically charged solutions is $F_{23}=q \sin\theta$,
therefore, the electromagnetic scalar, now has the form of
\begin{equation}
    F=\frac{q^2}{2 r^4}.
\end{equation}

Let us consider the following Lagrangian for our purposes~\cite{Singh:2022xgi}
\begin{equation}
     {\cal L}(F)=F\exp \left(-\frac{k \sqrt[4]{2 F q^2}}{q}\right), 
\end{equation}
where $k=q^2/2M$.

Using these quantities, we can therefore write the following line element from the equations of motion
\begin{eqnarray}
    ds^{2}&=&\left(1-\frac{2M}{r}e^{-\frac{q^{2}}{2Mr}}\right)dt^{2}-\left(1-\frac{2M}{r}e^{-\frac{q^{2}}{2Mr}}\right)^{-1}dr^{2}
	\nonumber \\    
 &&   -r^2\left(d\theta^{2}+\sin^{2}\theta d\phi^{2}\right)\,.
 \label{ds_Culetu}
\end{eqnarray}
This metric was also considered in \cite{Simpson:2019mud}, which was interpreted as asymptotically Minkowski core of a regular black hole solution.

The metric functions of Culetu's model are thus:
\begin{equation}
    A(r)=\frac{1}{B(r)}=1-\frac{2 M }{r}e^{ -\frac{q^2}{2 M r}},\quad C(r)=r^2\label{fm_Culetu}
\end{equation}
where $M$ and $q$ are, respectively, the mass of the black hole and the electric charge.

Next, we use the metric functions of the Culetu model~\eqref{fm_Culetu} to study the radius of the photon sphere. This can be achieved, for example, by the Lagrangian formalism. However, when it comes to GR  coupled with NLED, a more appropriate description is required to obtain the geodesics arising from this and other formulations. This development is achieved through the effective metric. We will then demonstrate the procedure by which we obtain geodesics, for example, for test particles in theories formulated with NLED.

\subsubsection{Effective metric}

However, since we have obtained solutions to the Einstein equations coupled with NLED described by the line element~\eqref{ds_Culetu}, we present the correct formulation for the geodesics. Indeed, the light rays now follow a geodesic described by the effective metric~\cite{Novello:1999pg,Toshmatov:2021fgm}
\begin{equation}
     g_{\rm eff}^{\mu\nu}={\cal L}_{F}g^{\mu\nu}-{\cal L}_{FF}F_{\sigma}^{\phantom{\sigma}\mu}F^{\sigma\nu}.\label{g_efet}
 \end{equation}
 We therefore have two line elements, depending on whether the electric or the magnetic charge plays a role.

The line element for the effective metric, if we consider the electric charge, is therefore given by 
\begin{eqnarray}
    ds^{2}&=&\frac{A(r)}{{\cal L}_{F}+2F{\cal L}_{FF}}dt^{2}-\frac{1}{B(r)\left({\cal L}_{F}+2F{\cal L}_{FF}\right)}dr^{2}
	\nonumber \\    
    && -\frac{r^{2}}{{\cal L}_{F}}\left(d\theta^{2}+\sin^{2}\theta d\phi^{2}\right).\label{ds_eletric}
\end{eqnarray}

On the other hand, the metric is expressed as follows when the magnetic charge is taken into account 
\begin{eqnarray}
    ds^{2}&=&\frac{A(r)}{{\cal L}_{F}}dt^{2}-\frac{1}{B(r){\cal L}_{F}}dr^{2}
	\nonumber \\    
   && -\frac{r^2}{{\cal L}_{F}+2F{\cal L}_{FF}}\left(d\theta^{2}+\sin^{2}\theta d\phi^{2}\right).\label{ds_mag}
\end{eqnarray}

However, since the metric function of the Culetu model results from the coupling of GR with an electromagnetic matter source from NLED, we now have that the particles describe a trajectory by an effective metric~\eqref{g_efet}.

Therefore, using the metric functions~\eqref{fm_Culetu} in the effective~\eqref{g_efet} metric, we get
\begin{subequations}\label{met_efet_Culetu}
    \begin{eqnarray}
  \bar{A}(r) &=& \frac{f(r)}{{\cal L}_{F}+2F{\cal L}_{FF}}= \frac{e^{-\frac{q^2}{m r}} }{32 m^2 r^3}\left(r e^{\frac{q^2}{2 m r}}-2 m\right) \times
	\nonumber \\  
  && \times \left(32 m^2 r^2-14 m q^2 r+q^4\right),
\\
    \bar{B}(r)&=& \frac{1}{f(r)\left({\cal L}_{F}+2F{\cal L}_{FF}\right)}
	\nonumber \\    
    &=& -\frac{32 m^2 r^2-14 m q^2 r+q^4}{64 m^3 r-32 m^2 r^2 e^{\frac{q^2}{2 m r}}},
    \\
    \bar{C}(r)&=& \frac{r^2}{{\cal L}_{F}}=-\frac{r e^{-\frac{q^2}{2 m r}} }{8 m}\left(q^2-8 m r\right).
    \end{eqnarray}
\end{subequations}
In the next subsections we analyze our results using the expressions~\eqref{met_efet_Culetu}.

\subsubsection{Geodesic formalism for the Culetu model}

Now, we apply the results described by the expressions~\eqref{met_efet_Culetu} to the geodesic equation~\eqref{eq_geo}. Thus, we obtain the form of Eq.~\eqref{R} for this case, given by:
\begin{eqnarray}
  &&   {\cal R}(r)=\frac{32M^{2}}{\left(32M^{2}r^{2}-14Mq^{2}r+q^{4}\right)^{2}}\Bigl[ 32{\cal E}^{2}M^{2}r^{4}e^{\frac{q^{2}}{Mr}}
	\nonumber \\     
   &&  -\frac{8L^{2}Me^{\frac{q^{2}}{Mr}}\left(2Me^{-\frac{q^{2}}{2Mr}}-r\right)\left(32M^{2}r^{2}-14Mq^{2}r+q^{4}\right)}{q^{2}-8Mr}\Biggr.
     	\nonumber\\
    &&\Biggl.+m^{2}r\left(2M-re^{\frac{q^{2}}{2Mr}}\right)\left(32M^{2}r^{2}-14Mq^{2}r+q^{4}\right)\Bigr].\label{R_Culetu}
\end{eqnarray}


On the other hand, if we insert ${\cal R}(r)=0$, i.e.  condition~\eqref{R0}, into Eq.~\eqref{R_Culetu}, we obtain ${\cal E}$. Consequently, it is possible to obtain $L$ by applying the condition~\eqref{dR0}, $\partial{\cal R}/\partial r=0$. Their explicit forms are described below:
\begin{eqnarray}
{\cal E} =\frac{ \sqrt{M-\frac{1}{2} r e^{\frac{q^2}{2 M r}}} \sqrt{32 M^2 r^2-14 M q^2 r+q^4} }{4 M r^2 \,e^{\frac{q^2}{2 M r}}\,\sqrt{q^2-8 M r}} \times
	\nonumber \\
	\times \sqrt{8 L^2 M e^{\frac{q^2}{2 M r}}+m^2 r \left(8 M r-q^2\right)}\,,
\end{eqnarray}
\begin{widetext}
\begin{eqnarray}
       L=m \sqrt{r}\left(q^{2}-8Mr\right)\left[q^{2}re^{\frac{q^{2}}{2Mr}}\left(60M^{2}r^{2}-18Mq^{2}r+q^{4}\right)+4M\left(4Mr-q^{2}\right)\left(8M^{2}r^{2}-13Mq^{2}r+q^{4}\right)\right]^{\frac{1}{2}}\times
       \nonumber\\
\times\sqrt{32M^{2}r^{2}-14Mq^{2}r+q^{4}}\Big/ \left(4\Biggl\{ M^2 e^{\frac{q^2}{2 M r}} \left(32 M^2 r^2-14 M q^2 r+q^4\right) \Bigr[-1536 M^4 r^4+1280 M^3 q^2 r^3 \Bigr. \Biggr.\right.\nonumber\\
\left.\Bigl. \Biggl.
-308 M^2 q^4 r^2-q^8+30 M q^6 r+
r^2 e^{\frac{q^2}{2 M r}} \left(512 M^3 r^3-368 M^2 q^2 r^2+60 M q^4 r-3 q^6\right)\Bigl] \Biggl\}^{\frac{1}{2}}\right)
.\label{L_Culetu}
\end{eqnarray}
\end{widetext}

Without a loss of generality, we can choose a value for the charge to obtain the radius of the photon sphere, for instance, from Eq.~\eqref{rps} for $q=0.5$, we obtain
\begin{equation}
    r_{ps}=2.842M.\label{rps_Culetu}
\end{equation}

Next, we calculate the $r_{ps}$ using the photon surface method.

\subsubsection{Photon surfaces in Culetu's model} 

We now determine $r_{ps}$ using the massive particle surface method. Taking into account the metric functions~\eqref{met_efet_Culetu} applied to Eq.~\eqref{phot_spher}, with the choice of charge being $q=0.5$, we obtain the following result:
\begin{equation}
    r_{ps}=2.842M.\label{rps2_Culetu}
\end{equation}
Note that the result for the radius of the photon sphere from the photon surface formalism, Eq.~\eqref{rps2_Culetu}, is the same as the result for the radius of the photon sphere from the geodesic formalism, Eq.~\eqref{rps_Culetu}.

\subsubsection{Massive particle surface for the Culetu model}

Let us now calculate the radius of the photon sphere by applying the formalism of the mass particle surface.

First, we obtain the electromagnetic tensor for this description, which is given by:
\begin{equation}
    F_{10}=\frac{q e^{-\frac{q^2}{2 M r}} \left(q^2-8 M r\right)}{8 M r^3}.\label{F10_CG}
\end{equation}
Integrating the above result, Eq.~\eqref{F10_CG}, with respect to the radial coordinate $r$, we obtain
\begin{equation}
    A_0=\frac{e^{-\frac{q^2}{2 M r}} \left(q^2-6 M r\right)}{4 q r}.
\end{equation}

Equation~\eqref{F_scalar} yields the following result for ${\cal F}$:
\begin{equation}
   {\cal F} =\frac{e^{-\frac{q^{2}}{2Mr}}\left(q^{3}-8Mqr\right)}{\sqrt{2}r^{3}}\left(\frac{r^{2}e^{\frac{q^{2}}{2Mr}}-2Mr}{32M^{2}r^{2}-14Mq^{2}r+q^{4}}\right)^{1/2}\,.
\end{equation}

Using the metric functions~\eqref{met_efet_Culetu}, the scalar of the second form, Eq.~\eqref{chi}, is given by
\begin{eqnarray}
   && \chi=-\frac{1}{r\left(8Mr-q^{2}\right)\left(32M^{2}r^{2}-14Mq^{2}r+q^{4}\right)^{2}} \times
	\nonumber \\   
  && \times \Big[4M\big(768M^{4}r^{4}+256M^{3}q^{2}r^{3}-280M^{2}q^{4}r^{2} +45Mq^{6}r
   \nonumber\\
&&-2q^{8}\big)
+re^{\frac{q^{2}}{2Mr}}\big(-2048M^{4}r^{4}+32M^{3}q^{2}r^{3} +372M^{2}q^{4}r^{2}
	\nonumber \\
&& -66Mq^{6}r+3q^{8}\big)\Big]\,\sqrt{\frac{2(32M^{2}r^{2}-14Mq^{2}r+q^{4})}{r\left(re^{\frac{q^{2}}{2Mr}}-2M\right)}}.
\end{eqnarray}
Consequently, the average curvature, as described by Eq.~\eqref{chi_afim}, is now expressed as:
\begin{equation}
   \chi_\lambda=\sqrt{\frac{e^{-\frac{q^2}{2 M r}} \left(r e^{\frac{q^2}{2 M r}}-2 M\right) \left(32 M^2 r^2-14 M q^2 r+q^4\right)}{M r^4 \left(8 M r-q^2\right)}}
\end{equation}
Now we can express $K$ with the help of these results:
\begin{widetext}
\begin{eqnarray}
    K &=& -\frac{2\sqrt{2}}{r\left(8Mr-q^{2}\right)\left(32M^{2}r^{2}-14Mq^{2}r+q^{4}\right)^{2}}\bigg[4M\left(768M^{4}r^{4}+256M^{3}q^{2}r^{3}-280M^{2}q^{4}r^{2}+45Mq^{6}r-2q^{8}\right)
    \nonumber\\
  && \hspace{1cm}  \left.+re^{\frac{q^{2}}{2Mr}}\left(-2048M^{4}r^{4}+32M^{3}q^{2}r^{3}+372M^{2}q^{4}r^{2}-66Mq^{6}r+3q^{8}\right)\right]\sqrt{\frac{32M^{2}r^{2}-14Mq^{2}r+q^{4}}{r\left(re^{\frac{q^{2}}{2Mr}}-2M\right)}}
   \nonumber\\
  &&  \hspace{2.5cm}  -3\sqrt{\frac{e^{-\frac{q^2}{2 M r}} \left(r e^{\frac{q^2}{2 M r}}-2 M\right) \left(32 M^2 r^2-14 M q^2 r+q^4\right)}{M r^4 \left(8 M r-q^2\right)}}.
\end{eqnarray}
\end{widetext}


With these results, we have also derived the energy ${\cal E}$ of the test particle, however, since the expression is quite extensive, we will not state it explicitly. Thus, if we consider $Q=0$ and $q=0.5$ in this energy for the condition $1/{\cal E}^2=0$, we find that the radius of the photon sphere is given by:
\begin{equation}
    r_{ps}=3.12783M.\label{rps3_Culetu}
\end{equation}

We thus note that the result of the method of massive particle surfaces, as in Eq. \eqref{rps3_Culetu}, does not agree with the other two results we previously obtained when using the geodesic and photon surface formalism, as shown in Eqs.~\eqref{rps_Culetu} and~\eqref{rps2_Culetu}, respectively. This indicates a lack of agreement between the three formalisms and points to an apparent inconsistency in the use of non-linear electrodynamics metric functions, i.e. for the effective~\eqref{g_efet} metric.  We also verify that this discrepancy occurs in the case of a regular black hole solution of Bardeen described by NLED.

\section{Conclusion}\label{sec6}

In this work, we explored three distinct methods to determine, for instance, the radius of the photon sphere, denoted as $r_{ps}$, through the following approaches: geodesic, photon surface, and massive particle surface. In our results, we employed two gravitational models. The first is described by conformal gravity, briefly highlighting its specific characteristics, while the second is the Culetu model, derived from NLED.

In the geodesic formalism, the orbits of the particles are determined by solving the geodesic equations, which is initially achieved by the Lagrangian. In this context, the radius of the photon sphere is calculated taking into account the null geodesics describing the photon orbits. On the other hand, the photon surface proposed by Ellis et al. offers a more comprehensive formulation. In this formalism, the null geodesics, which were originally tangent to the surface, remain trapped in this hypersurface. The massive particle surface method extends this analysis even further, allowing not only photons but also charge massive particles to be considered. This formalism provides a more general description and its results should agree with previous methods when applied to photons.

In our work, our main objective was to highlight some slight inconsistencies between the geodesic and photon surface methods with the massive particle surface method.
We have obtained results that are inconsistent with the massive particle surface method.
In Section \ref{sec5}, we have used two gravitational models that we apply to the formalisms studied in this manuscript. We found that the massive particle surface method needs to be generalized to become more comprehensive and to consider a wider range of cases. In the model of conformal gravity described by metric functions with $g_{00}\neq -g_{11}^{-1}$, we have found that the result for $r_{ps}$, as in Eq.~\eqref{rps3_CG}, is not consistent with the results of the other two formalisms, Eqs.~\eqref{rps_CG} and~\eqref{rps2_CG}. We have also found that the values for $r_{\rm ISCO}$ determined by the  photons surfaces method and massive particle surfaces method do not agree, as described in Eqs.~\eqref{risco_CG} and ~\eqref{risco2_CG}.

We also examined the results of the Culetu model for these three methods. The Culetu model was developed by coupling GR  with the matter source of NLED and represents a static and spherically symmetric line element Eq.~\eqref{ds_Culetu}. Thus, when dealing with a model developed using NLED, one has to consider a correction of the line element, for example to study geodesics. This is done by the effective metric Eq.~\eqref{g_efet}. We have applied the metric functions of this model Eq.~\eqref{met_efet_Culetu}, which now result from the correction of the effective metric, to obtain the photon sphere radius ($r_{ps}$) in the three formalisms. We found that the values for the photon sphere radius obtained from the three formalisms do not agree, as can be seen from Eqs.~\eqref{rps_Culetu}, ~\eqref{rps2_Culetu} and ~\eqref{rps3_Culetu}. For clarity, we have created a table in which we summarize these discrepancies, see Table~\ref{tab}.

\begin{table*}
\resizebox{18cm}{!} 
{
\begin{tabular}{|>{\centering}p{2cm}|c|c|c|c|c|}
 \cline{2-6} \cline{3-6} \cline{4-6} \cline{5-6} \cline{6-6} 
\multicolumn{1}{c|}{} & \multicolumn{5}{c|}{\large{\textbf{Formalism}}}\tabularnewline
\hline
\textbf{Metric function} & \multicolumn{2}{c|}{\textbf{Geodesic}} & \textbf{Photon surface}
 & \multicolumn{2}{c|}{\textbf{Massive particle surface}}\tabularnewline
\hline 
Reissner-Nordström & $r_{ps}=\frac{1}{2}\left(\sqrt{9M^{2}-8q^{2}}+3M\right)$ & $r_{\rm ISCO}=5.939M$ & $r_{ps}=\frac{1}{2}\left(\sqrt{9M^{2}-8q^{2}}+3M\right)$ & $r_{ps}=\frac{1}{2}\left(\sqrt{9M^{2}-8q^{2}}+3M\right)$ & $r_{\rm ISCO}=5.939M$\tabularnewline
\hline 
Conformal gravity & $r_{ps}=3M$ & $r_{\rm ISCO}=5.975M$ & $r_{ps}=3M$ & $r_{ps}=2.925M$ & $r_{\rm ISCO}=5.702M$\tabularnewline
\hline 
Culetu (effective metric) &  $r_{ps}=2.842M$ & \resizebox{0.6cm}{0.6cm}{---} & $r_{ps}=2.842M$ & $r_{ps}=3.127M$ & \resizebox{0.6cm}{0.6cm}{---} \tabularnewline
\hline 
\end{tabular}
}
\caption{This table summarizes the results obtained for $r_{ps}$ and $r_{\rm ISCO}$ in the Reissner-Nordström, Conformal gravity and Culetu's model with effective metrics.}
\label{tab}
\end{table*}

The success of these formalisms is remarkable, and they have, for example, played an important role in understanding the different approaches to the strong-field regime. Given the contradictions  in the results that have emerged in this study, it is not our aim to generally point out shortcomings in the existing methods or to propose a new method. Rather, we would like to point out that these formalisms can still be extended to cover an even wider range of gravitational models.


To conclude this manuscript, we comment on a general result for the radius of the photon sphere ($r_{ps}$) and the radius ($r_{\rm ISCO}$) through the formalism of massive particle surfaces. To illustrate this, consider the following metric
\begin{equation}
     ds^{2}=A(r)dt^{2}-B(r)dr^{2}-C(r)\left(d\theta^{2}+\sin^2\theta d\phi^{2}\right).\label{metric_concl}
\end{equation}
We emphasize that we will focus on the case of a neutral test particle to derive the expressions that are presented below.

Using the metric~\eqref{metric_concl} in the relevant quantities that were presented in Section~\ref{sec4}, we find that the energy of the test particle in its most general form is described by
\begin{equation}
     {\cal E}=\sqrt{\frac{2m^2 A^{2}(r)\sqrt{B(r)}}{6A(r)\sqrt{B(r)}-\sqrt{\frac{C(r)}{A(r)}}A'(r)-2\sqrt{\frac{A(r)}{C(r)}}C'(r)}}.\label{Econcl}
\end{equation}


If we apply the metric functions described by the Schwarzschild model in Eq.~\eqref{Econcl} and then take into account that the energy of the particle diverges, we get the same result for the photon sphere in the geodesic and photon surface methods, that is, $r_{ps}=3M$, as shown in e.g. Eq.~\eqref{rpsi}.

Furthermore, if we take the derivative of the energy of the particle described by Eq.~\eqref{Econcl} and apply the condition $d{\cal E}/dr=0$ to its result, we get the following expression:
\begin{widetext}
\begin{align}
  &  A(r)C(r)\left[12B(r)^{3/2}C(r)\sqrt{\frac{A(r)}{C(r)}}A'(r)+B(r)\Big(2c(r)A''(r)-5A'(r)C'(r)\Big)-C(r)A'(r)B'(r)\right]\nonumber\\
   & -5B(r)C^{2}(r)\big(A'(r)\big)^{2}-2A^{2}(r)\left[C(r)\big(B'(r)C'(r)-2B(r)C''(r)\big)+B(r)\big(C'(r)\big)^{2}\right]=0.\label{dr2conc}
\end{align}
\end{widetext}

This is a general expression with which we can calculate the radius $r_{\rm ISCO}$. When we apply the metric functions of the Schwarzschild model to this expression, we get the same result as the geodesic method, i.e. $r_{\rm ISCO}=6M$. It is also worth mentioning that this expression, Eq.~\eqref{dr2conc}, differs significantly from the equation obtained by the geodesic formalism Eq.~\eqref{dR2_geod}. Moreover, its structure shows a dependence on the metric function $B(r)$ to determine the ($r_{\rm ISCO}$).

In general, by imposing the condition $1/{\cal E}^2=0$ on the energy Eq.~\eqref{Econcl}, we observe that the method of massive particle surfaces results in the following expression, allowing us to obtain the radius of the photon sphere
\begin{equation}
    6\sqrt{A^{3}(r)B(r)C(r)}-\Big(C(r)A'(r)+2A(r)C'(r)\Big)=0.\label{SPM_phot_spher}
\end{equation}

We see in Eq.~\eqref{SPM_phot_spher} that it differs significantly from the result obtained with the photon surface method, as shown in Eq.~\eqref{phot_spher}. From this we can conclude that when applying the Schwarzschild model, the results for the radius of the photon sphere and $r_{\rm ISCO}$ the three formalisms have the same values. In other words, we observe that this method leads to the usual photon sphere for some special cases. However, when we consider a general metric, we find that the expressions we found for these quantities, $r_{ps}$ and $r_{\rm ISCO}$, using the massive particle formalism are not equivalent to other methods and exhibit a dependence on the metric function $B(r)$, as we can see in Eqs.~\eqref{dr2conc} and~\eqref{SPM_phot_spher}.

\acknowledgments

MER  thanks Conselho Nacional de Desenvolvimento Cient\'ifico e Tecnol\'ogico - CNPq, Brazil, for partial financial support. This study was financed in part by the Coordena\c{c}\~{a}o de Aperfei\c{c}oamento de Pessoal de N\'{i}vel Superior - Brasil (CAPES) - Finance Code 001.
FSNL acknowledges support from the Funda\c{c}\~{a}o para a Ci\^{e}ncia e a Tecnologia (FCT) Scientific Employment Stimulus contract with reference CEECINST/00032/2018.  FSNL and LFDS acknowledge funding through the research grants UIDB/04434/2020, UIDP/04434/2020 and PTDC/FIS-AST/0054/2021.


\end{document}